# Initialization-free ultralow-power programmable spin-orbit torque logic devices enabled by chirally asymmetric switching

Qianbiao Liu, Lijun Zhu*

*State Key Laboratory of Semiconductor Physics and Chip Technologies, Institute of Semiconductors, Chinese Academy of Sciences, Beijing 100083, China*

*ljzhu@semi.ac.cn

**Abstract: Spin logic is of great interest for the development of high-performance, non-Von Neumann artificial intelligence chips. Of particular advantage is the programmable spin-orbit torque logic that can achieve the complete set of the 16 Boolean logic operations within a single device in an initialization-free, low-power, and scalable manner, which is, however, a challenge despite remarkable efforts over more than two decades. Here, utilizing chirally asymmetric switching, we demonstrate an initialization-free, low-power, programmable spin-orbit torque logic device that is capable of the complete set of 16 Boolean logic operations within a single device with only three inputs and ultralow power of < 1 fJ/bit. We also propose the first initialization-free, all-electrical spin-based cascading computing devices, including a half adder, a three-level full adder, a two-level full adder, and a Transfer/NOT selector. These compact and scalable computing devices are enabled by the chirally asymmetric spin-orbit torque switching of a Ta/FeCoB bilayer with a significant perpendicular Dzyaloshinskii-Moriya interaction field. These results pave an intriguing way for the development of next-generation high-performance large-scale in-memory computing chips based on chirally asymmetric spin-orbit torque switching.**

**Introduction.** The rapid growth of artificial intelligence technologies creates computing demand that is doubling every two months. However, the traditional von Neumann computing (Fig. 1a) shows stagnation due to the limited energy-efficiency and speed of the charge-based logic and memory and due to the additional time delay, power, and data bandwidth shortage induced by the logic-memory interconnection. To overcome these limitations, considerable efforts have been made to develop fast, energy-efficient, dense in-memory computing spin logic devices over the two decades [1-31]. While all the gate functions can be constructed by cascading the three basic gates of AND, OR, and NOT (Fig. 1a), the advantage is degraded due to the considerable cost of circuit complications, power increase, and integration density loss. In contrast, it is of particular promise to develop programmable spin-orbit torque (SOT) logic that can achieve the complete set of 16 Boolean logic operations (i.e., AND, OR, NOT, NOR, NAND, XOR, XNOR, Transfer, Null, Identity, inverse implication (NIMP), implication (IMP), reverse inverse implication (RNIMP), and reverse implication (RIMP)) within a single device in an initialization-free, low-power, and scalable manner. So far, this has remained a major challenge.

Here, we for the first time demonstrate an initialization-free, scalable, ultralow-power programmable Ta/FeCoB SOT logic device that has the 16 complete Boolean logic gate functions and the potential of wafer-scale integration with CMOS circuits and magnetic tunnel junctions (MTJs), all electrical operations, and compact construction of cascading computing devices (e.g., half adder, full adder). Such excellent programmability is enabled by the chirally asymmetric SOT switching due to the perpendicular effective magnetic field ($H_{\rm DMI}^{z}$) arising from the interplay of the Dzyaloshinskii-Moriya interaction (DMI) and anisotropy fluctuations (Fig. 1b)[31].

**Device characterizations.** A Ta (5nm)/FeCoB (1.3nm) bilayer (FeCoB=$Fe_{60}Co_{20}B_{20}$) with perpendicular magnetic anisotropy was sputter-deposited on an oxidized Si wafer and protected from oxidation by a MgO (1.6nm)/$TaO_x$ (3nm) bilayer. The sample is then patterned into 5 μm-wide Hall-cross devices (Fig. 1a) with two current inputs ($A$ and $B$). The FeCoB, which allows for sensitive readout by MTJs with large tunneling magnetoresistance (TMR), functions as the data bit with the magnetization pointing up (down) as the "1" ("0") state. The Ta layer generates spin current via the spin Hall effect for the SOT manipulation of the data bit states. The state of the FeCoB bit is read via the anomalous Hall resistance ($R_H$).

A key knob for programming the logic in this work is the high-degree asymmetry of the switching currents. As shown in Fig. 1c, the Ta/FeCoB device requires a total current ($A + B$) of 3.1 mA (-1.2 mA) for upward switching and -1.2 mA (3.1 mA) for downward switching under an in-plane magnetic field ($H_x$) of 600 Oe (-600Oe), which suggests that $H_{\rm DMI}^{z}$ favors "0" ("1") states. In Fig. 1d, we plot the total switching current of the Ta/FeCoB device ($H_x$ = 600 Oe) as a function of pulse width, fit of which to the equation [32] of $I_c = I_{c0}[1- \Delta^{-1}\ln(\tau/\tau_0)]$ yields a thermal stability factor $\Delta$ of 20 (13) and critical switching current $I_{c0}$ of 3.7 (-2.4) mA for upward and downward switching ($\tau_0 \approx$ 1ns [32]). This result suggests an ultralow power of 0.85 (0.36) fJ/bit at 1 ns pulse width and an industrial-level current channel dimension of 50×100 nm$^2$.

As shown in Fig. 1e, the values of $H_{\rm DMI}^{z}$ for the Ta/FeCoB device are measured from the $H_x$-induced anomalous Hall resistance loop shift, i.e., $H_{\rm DMI}^{z} = (H_\uparrow + H_\downarrow)/2$, where $H_\uparrow$ and $H_\downarrow$ denote the upward and downward switching fields. Here, only a small in-plane current (0.1 mA) was applied for detecting the anomalous Hall resistance without generating any significant SOT field (<0.4 Oe). As plotted in Fig. 1f the coercivity ($H_c = |H_\uparrow - H_\downarrow|/2$) varies symmetrically with $H_x$ while $H_{\rm DMI}^{z}$ is an anti-symmetric function of $H_x$, which are consistent with other heavy metal/ferromagnet devices [31,33,34].

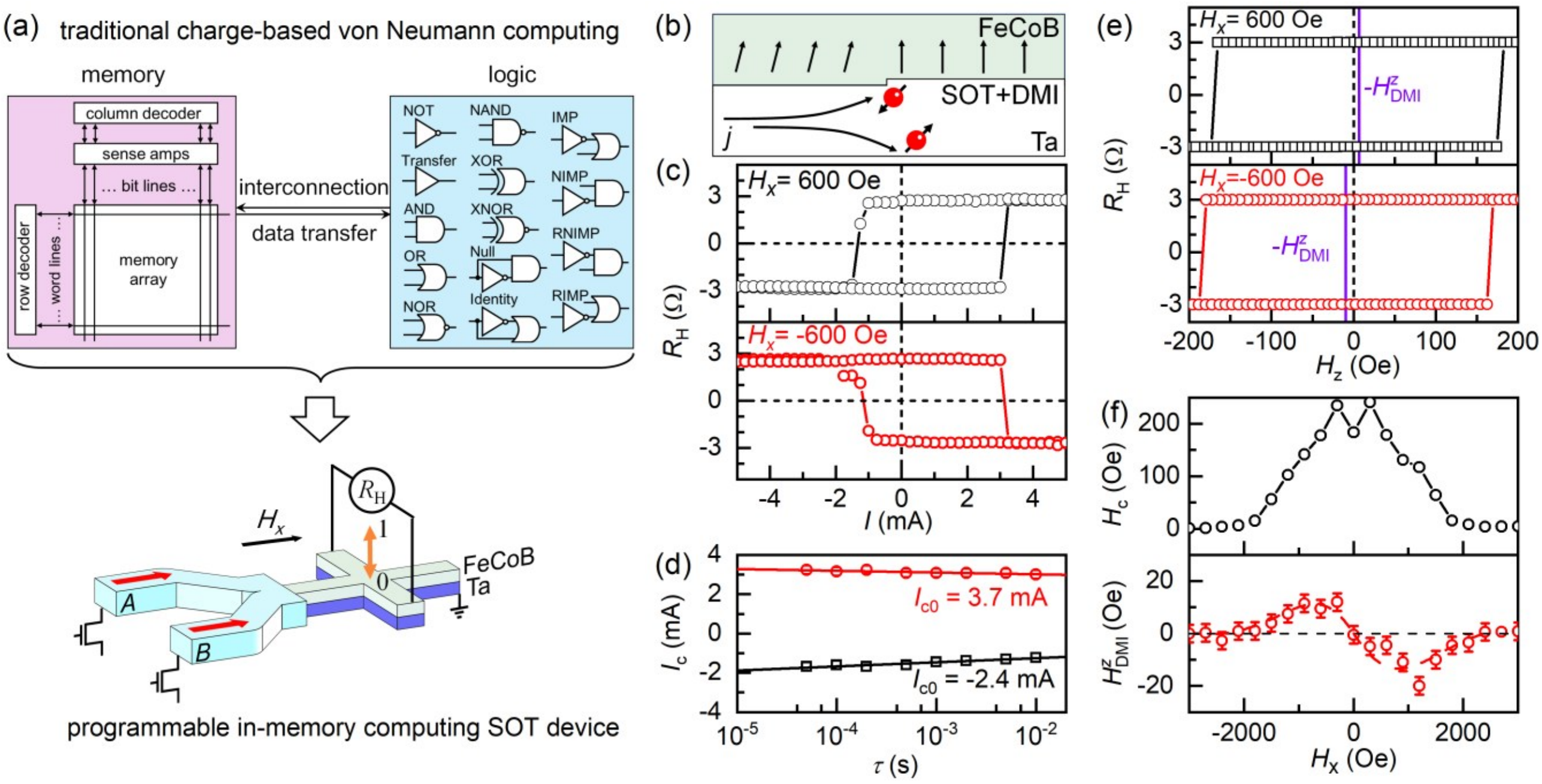

**Fig. 1| Switching current asymmetry for logic programming.** (a) Schematic of the traditional charge-based von Neumann computing vs the programmable SOT logic device with current inputs A and B, in-plane magnetic field ($H_x$), and Hall detection of "0" and "1" states. (b) Asymmetric switching currents ($H_x = \pm$ 600 Oe). (c) Total switching current ($H_x$ = 600 Oe) plotted as a function of write pulse duration ($\tau$), with two straight lines representing the best logarithmic fits of the data. (d) Hall resistance hysteresis driven by the perpendicular magnetic field ($H_z$) ($H_x$ = ±600 Oe). (e) Dependences on $H_x$ of the coercivity ($H_c$) and the chiral perpendicular field ($H_{\rm DMI}^{z}$).

**Initialization-free complete programmable logic gates.** We experimentally demonstrate all the 16 logic gate operations in a single programmable device by taking advantage of the asymmetric switching of the Ta/FeCoB device. We employ two currents (A, B) and one magnetic field ($H_x$) as the inputs. For all the logic operations, the magnetic field input is 600 Oe for "0" and -600 Oe for "1".

The device achieves the logic operations of AND, NAND, XOR, Transfer A(B), NOT A(B), Identity, and Null with -4 mA as "0" and 2 mA as "1" for the inputs A and B (Fig. 2a). Specifically, the device becomes an XOR gate for $B$ = "1" and outputs "0" only when $A$ = $H_x$ = "1" or "0" (Fig. 2b). In Fig. 2c the device becomes an AND (NAND) gate for $H_x$ = "0" ("1") and outputs "1" ("0") only if $A$ = $B$= "1". In Fig. 2d,e, for a fixed $B$ ($A$) at "1", the AND gate is reduced to Transfer-A (Transfer-B), while the NAND gate is reduced to NOT-A (NOT-B). When $A$ (or $B$) is fixed at "0", the AND gate becomes a Null gate, while the NAND gate becomes an Identity gate (Fig. 2f).

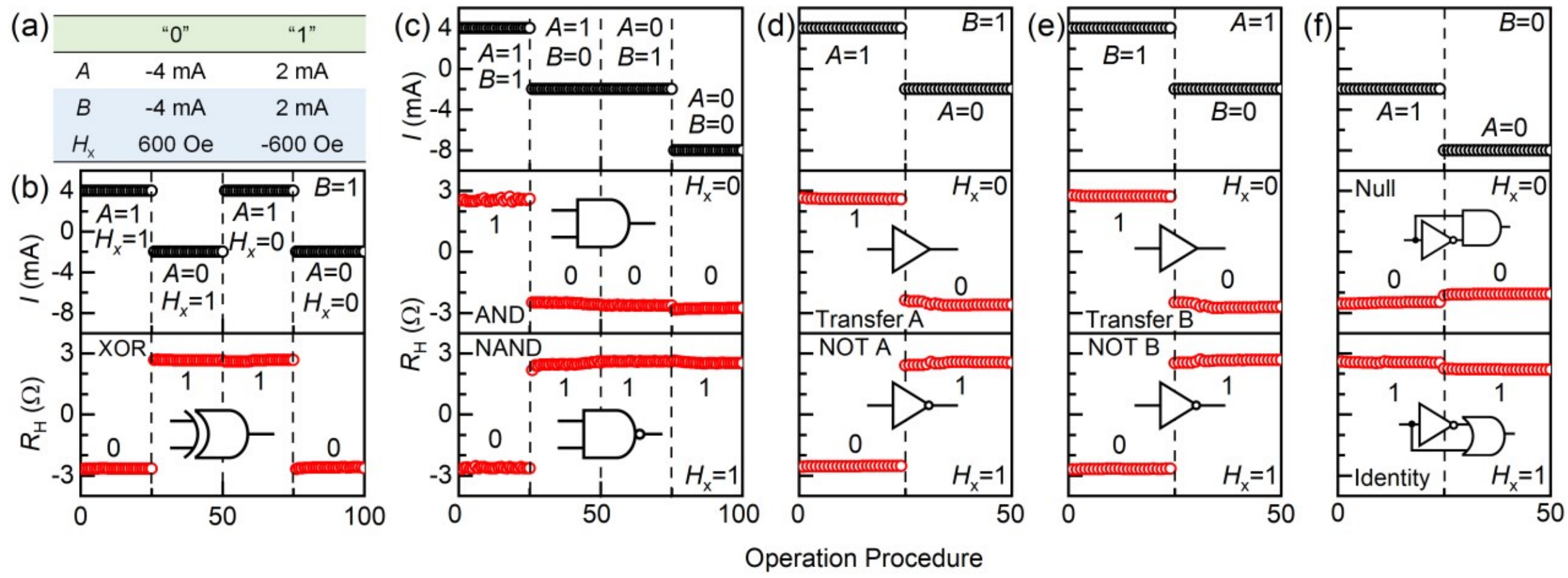

**Fig. 2| XOR, AND, NAND, Transfer A, NOT A, Transfer B, NOT B, Null, and Identity gates.** (a) Definition of the logic values of the input currents (A, B) and in-plane magnetic field ($H_x$). Current and Hall resistance for programmable operations of (b) XOR, (c) AND and NAND, (d) Transfer A and NOT A, (e) Transfer B and NOT B, (f) Null and Identity.

We show in Fig. 3a-c that the device can achieve OR, NOR, and XNOR operations with 2 mA as the input "0" and -4 mA as the input "1" for both $A$ and $B$. As shown in Fig. 3b, for $B$ fixed at "0", the device functions as an XNOR gate that returns "1" when $A$ = $H_x$ = "1" (or "0"). In Fig. 3c, the device becomes an OR gate that returns "1" except when $A$ =$B$ = "0" for $H_x$ fixed at "1" but a NOR gate that returns "0" unless $A$ = $B$ ="0" for $H_x$ fixed at "0".

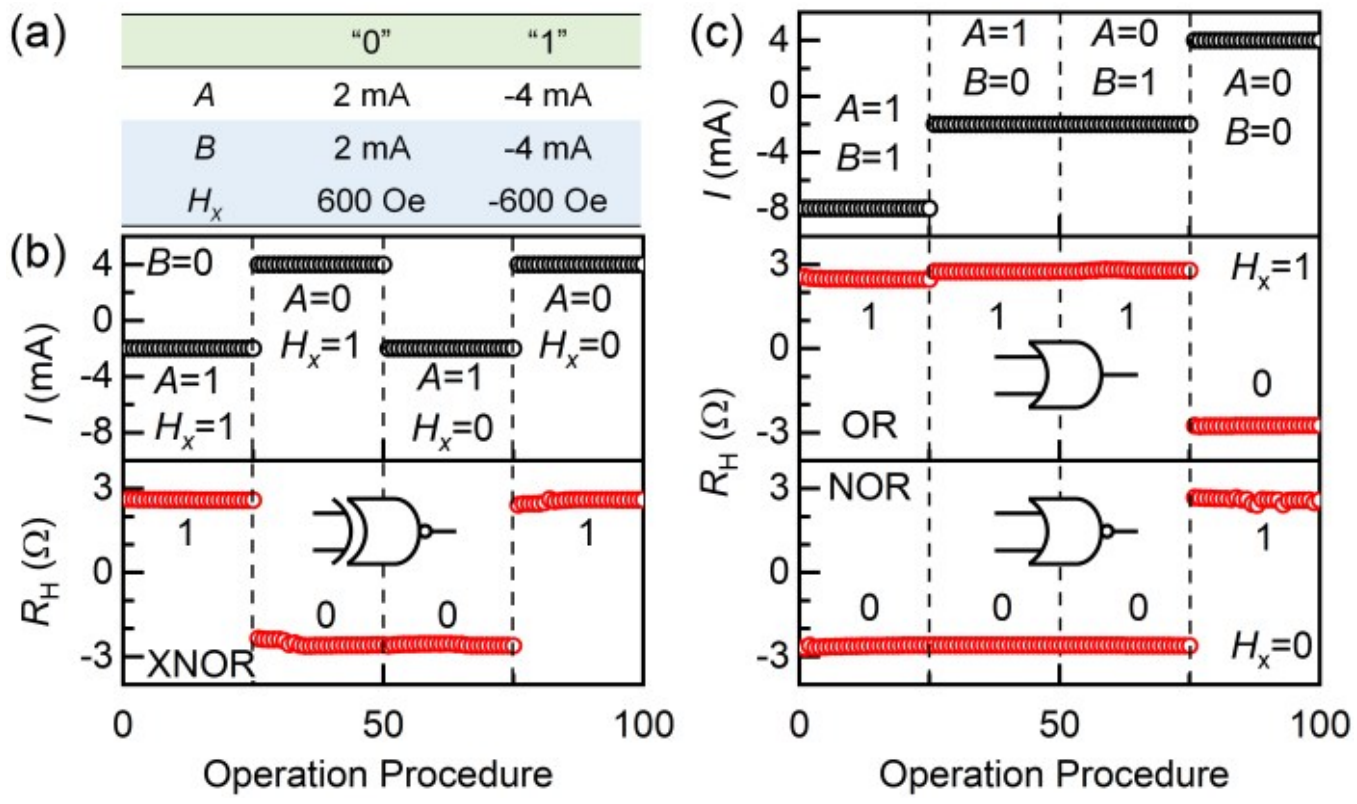

**Fig. 3| XNOR, OR, and NOR gates.** (a) Definition of the logic values of the input currents (A, B) and in-plane magnetic field ($H_x$). Current and Hall resistance for operations of (b) XNOR and (c) OR and NOR.

As shown in Fig. 4a, when we define -4 mA as "0" and 2 mA as "1" for $A$, and 2 mA as "0" and -4 mA as "1" for $B$, the device can achieve the functions of NIMP for $H_x$ fixed at "0" and IMP for $H_x$ fixed at "1".

IMP returns "0" only if $A$ = "0" and $B$ = "1", while NIMP returns "1" only if $A$ = "1" and $B$ = "0". In Fig. 4b, when we define 2 mA as "0" and -4 mA as "1" for $A$, and -4 mA as "0" and 2 mA as "1" for $B$, the device can achieve the functions of RNIMP for $H_x$ fixed at "0" and RIMP for $H_x$ fixed at "1". RNIMP returns "1" only if $A$ = "0" and $B$ = "1", while RIMP returns "0" only if $A$ = "1" and $B$ = "0".

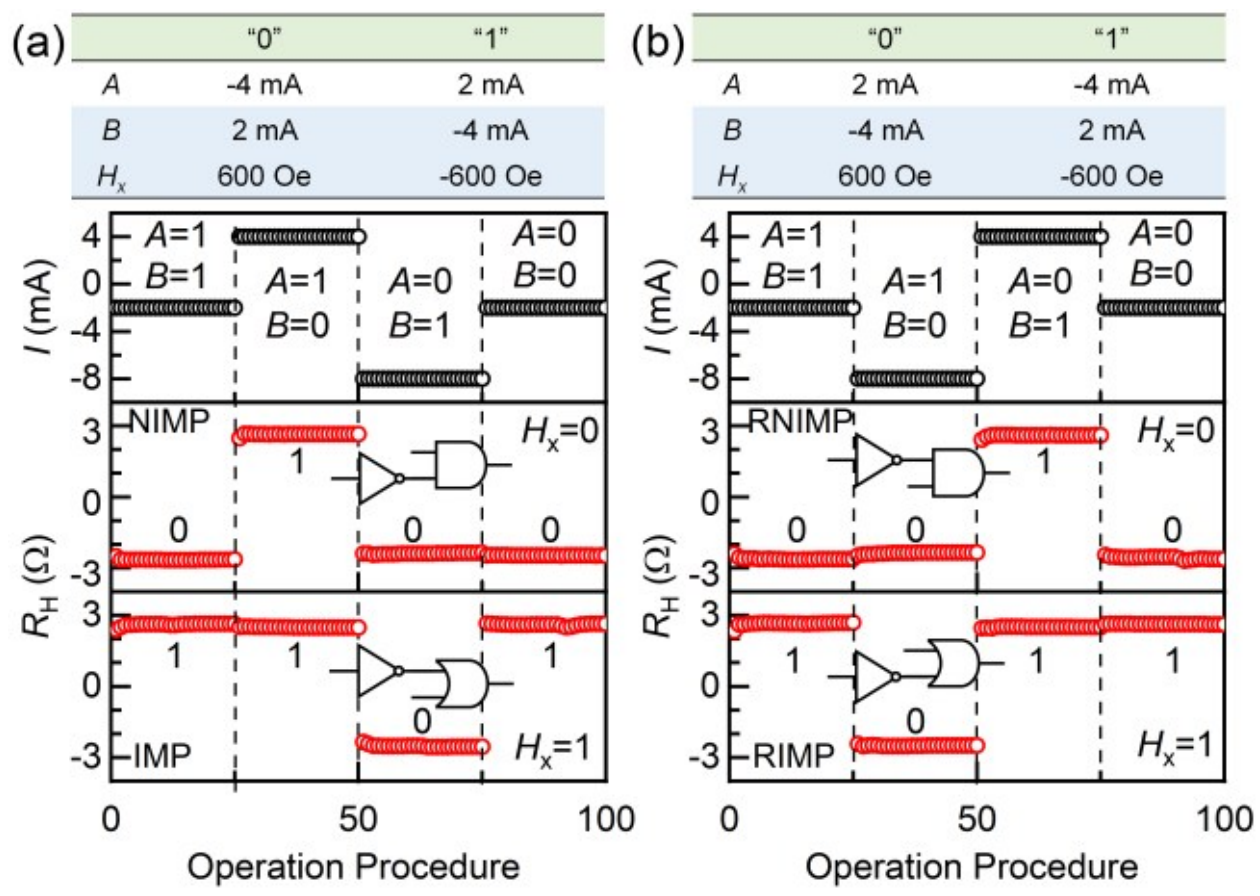

**Fig. 4| NIMP, IMP, RNIMP, and RIMP gates**. Input definitions and programmable operations of (a) NIMP, IMP and (b) RNIMP, RIMP.

These results represent the demonstration of the initialization-free complete 16 Boolean logic gate operations within a single device with only 3 inputs. Importantly, the operation of the programmable logic device in this work requires only two constant currents and a constant bias field magnitude. Such devices have the potential for field-free operation because the constant-magnitude in-plane magnetic field can be replaced by an on-chip in-plane nanomagnet switchable by the SOT. The simple architecture also allows for high-sensitivity readout using TMR of MTJs.

**Initialization-free, all-electrical cascading computing devices.** With the above programmable SOT gates (with write current of $I_0$), we propose compact, initialization-free, all-electrical, and scalable cascading computing devices of half adder, full adder, and selector. As shown in Fig. 5, we propose to introduce an on-chip in-plane magnet controllable by a spin Hall current channel to provide the in-plane field (i.e., $H_x$ in Figs. 1-4) for all-electrical operation of all the gates. Here, the in-plane magnet is supposed to have a switching current of less than $I_0$, which is usually the case as indicated by previous in-plane SOT-MTJ experiments [32,35]. The Hall-cross logic bit is also replaced by a MTJ (cascading operation of which is feasible, e.g., by cross-array [36]) for enhanced cascading and output performance.

As a basic computing device, a half adder adds two single binary bits (A, B) to produce a Sum and Carry but can't handle a carry-in. As schematically shown in Fig. 5a, the half adder cascades an AND gate and a XOR gate as Sum and Carry outputs and has two current inputs (A and B). With a switching current of $I_0$ as "1" for A and B, -(1+$\eta$) $I_0$ as "0" for A and B ($\eta$ is the magnitude ratio of the smaller and greater currents for upward and downward switching, see Fig. 1b,c), the half adder outputs Carry of "1" and Sum of "0" for $A$ = $B$ = "1" and Carry of "0" and Sum of "0" for $A$ = $B$ = "0". Otherwise, it outputs Carry of "0" and Sum of "1".

The more essential computing device for multi-bit addition is the full adder that adds three bits (A, B, and Carry-in C) to produce a Sum and Carry-out. In Fig. 5b we first propose a three-input full adder design by cascading two half adders (Fig. 5a) and one OR gate following the conventional three-level five-gate architecture. In Fig. 5c, we further propose a simplified two-level three-device full adder architecture. The Carry is a three-input SOT-MTJ device with a fixed in-plane magnet, while the sum is a three-input SOT-MTJ device with an in-plane magnet controlled by spin current channel with a Transfer/NOT selector (Fig. 5d). We define input current $I_0$ as "1" for A and B, $(1+\eta)I_0$ as "1" for C, $-(1+\eta)I_0$ as "0" for A and B, $-I_0$ as "0" for C. The full adder outputs Carry of "1" only for the case of more than one "1" inputs and Sum of "1" only for the case of an odd number of "1" inputs. As plotted in Fig. 5d, the selector includes a SOT-MTJ device with a current input C and an in-plane magnet controlled by a spin current channel with inputs A and B. For all three inputs, "1" is defined as $(1+\eta)I_0$ but "0" as $-I_0$. The selector functions as a NOT gate for $A$=$B$="0" and as a Transfer gate otherwise.

As compared in Fig. 5e, the two-level three-device full adder is the first initialization-free spintronic full adder. It is expected to greatly simplify the fabrication and improve the energy-efficiency and latency compared to other proposals (note that a domain wall full adder [13] consists of 15 NAND and 3 NOT gates).

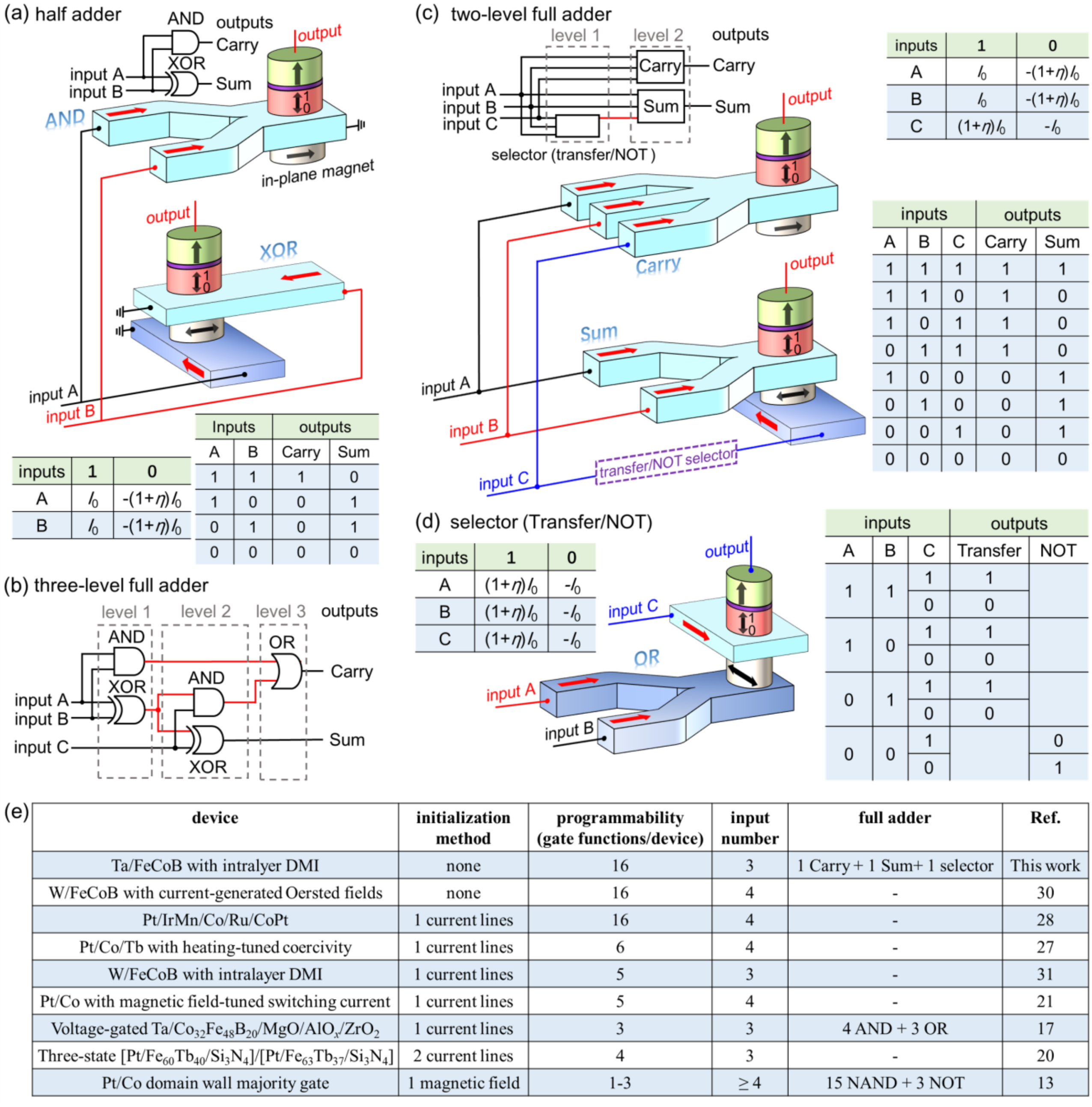

| inputs | 1 | 0 |
|---|---|---|
| A | $I_0$ | $-(1+\eta)I_0$ |
| B | $I_0$ | $-(1+\eta)I_0$ |

| Inputs | | outputs | |
|---|---|---|---|
| A | B | Carry | Sum |
| 1 | 1 | 1 | 0 |
| 1 | 0 | 0 | 1 |
| 0 | 1 | 0 | 1 |
| 0 | 0 | 0 | 0 |

| inputs | 1 | 0 |
|---|---|---|
| A | $I_0$ | $-(1+\eta)I_0$ |
| B | $I_0$ | $-(1+\eta)I_0$ |
| C | $(1+\eta)I_0$ | $-I_0$ |

| inputs | | | outputs | |
|---|---|---|---|---|
| A | B | C | Carry | Sum |
| 1 | 1 | 1 | 1 | 1 |
| 1 | 1 | 0 | 1 | 0 |
| 1 | 0 | 1 | 1 | 0 |
| 0 | 1 | 1 | 1 | 0 |
| 1 | 0 | 0 | 0 | 1 |
| 0 | 1 | 0 | 0 | 1 |
| 0 | 0 | 1 | 0 | 1 |
| 0 | 0 | 0 | 0 | 0 |

| inputs | 1 | 0 |
|---|---|---|
| A | $(1+\eta)I_0$ | $-I_0$ |
| B | $(1+\eta)I_0$ | $-I_0$ |
| C | $(1+\eta)I_0$ | $-I_0$ |

| inputs | | | outputs | |
|---|---|---|---|---|
| A | B | C | Transfer | NOT |
| 1 | 1 | 1 | 1 | |
| | | 0 | 0 | |
| 1 | 0 | 1 | 1 | |
| | | 0 | 0 | |
| 0 | 1 | 1 | 1 | |
| | | 0 | 0 | |
| 0 | 0 | 1 | | 0 |
| | | 0 | | 1 |

| device | initialization method | programmability (gate functions/device) | input number | full adder | Ref. |
|---|---|---|---|---|---|
| Ta/FeCoB with intralyer DMI | none | 16 | 3 | 1 Carry + 1 Sum+ 1 selector | This work |
| W/FeCoB with current-generated Oersted fields | none | 16 | 4 | - | 30 |
| Pt/IrMn/Co/Ru/CoPt | 1 current lines | 16 | 4 | - | 28 |
| Pt/Co/Tb with heating-tuned coercivity | 1 current lines | 6 | 4 | - | 27 |
| W/FeCoB with intralayer DMI | 1 current lines | 5 | 3 | - | 31 |
| Pt/Co with magnetic field-tuned switching current | 1 current lines | 5 | 4 | - | 21 |
| Voltage-gated $Ta/Co_{32}Fe_{48}B_{20}/MgO/AlO_x/ZrO_2$ | 1 current lines | 3 | 3 | 4 AND + 3 OR | 17 |
| Three-state $[Pt/Fe_{60}Tb_{40}/Si_3N_4]/[Pt/Fe_{63}Tb_{37}/Si_3N_4]$ | 2 current lines | 4 | 3 | - | 20 |
| Pt/Co domain wall majority gate | 1 magnetic field | 1-3 | ≥ 4 | 15 NAND + 3 NOT | 13 |

**Fig. 5| Proposal of computing devices based on the programmable gates.** (a) Half adder. (b) Three-level full adder, (c) Two-level full adder, (d) Transfer/NOT selector. The truth value table is for the spin Hall channel with a negative spin Hall ratio. (e) Comparison of logic device performance with other works.

**Conclusion.** Taking advantage of asymmetric spin-orbit torque switching of a Ta/FeCoB bilayer with a significant perpendicular DMI field, we have developed the first initialization-free programmable SOT logic device that is capable of the complete set of 16 Boolean logic operations within a single device with only three inputs and ultralow power of < 1 fJ/bit. Such programmable SOT logic devices can remarkably simplify the computing circuits compared to the traditional charge-based von Neumann architecture made of unprogrammable complex transistor logics and semiconductor capacitor memory (Fig. 1a) and other spintronic logic proposals (see comparison in Fig. 5e). Such logic devices are advantageous also for the potential of all-electrical operations, excellent cascadability, scalability, and integration compatibility with CMOS circuits and high-output MTJs. With the programmable SOT logic devices, we further propose the first initialization-free, all-electrical, and scalable computing devices, including a half adder, three-level full adder, two-level full adder, and a selector. These results pave an intriguing way for the development of next-generation high-performance large-scale in-memory computing AI chips based on chirally asymmetric SOT switching.

## REFERENCES

[1] D. A. Allwood, G. Xiong, M. D. Cooke, C. C. Faulkner, D. Atkinson, N. Vernier, and R. P. Cowburn, Submicrometer ferromagnetic NOT gate and shift register, Science 296, 2003 (2002).

[2] A. Ney, C. Pampuch, R. Koch, and K. H. Ploog, Programmable computing with a single magnetoresistive element, Nature 425, 48 (2003).

[3] H. Dery, P. Dalal, Ł. Cywiński, and L. J. Sham, Spin-based logic in semiconductors for reconfigurable large-scale circuits, Nature 447, 573 (2007).

[4] B. Behin-Aein, D. Datta, S. Salahuddin, and S. Datta, Proposal for an all-spin logic device with built-in memory, Nat. Nanotech. 5, 266 (2010).

[5] D. E. Nikonov, G. I. Bourianoff, and T. Ghani, Proposal of a spin torque majority gate logic, IEEE Electron Device Lett. 32, 1128 (2011).

[6] W. A. Borders, A. Z. Pervaiz, S. Fukami, K. Y. Camsari, H. Ohno, and S. Datta, Integer factorization using stochastic magnetic tunnel junctions, Nature 573, 390 (2019).

[7] J. Wunderlich, B.-G. Park, A. C. Irvine, L. P. Zârbo, E. Rozkotová, P. Nemec, V. Novák, J. Sinova, and T. Jungwirth, Spin Hall effect transistor, Science 330, 1801 (2010).

[8] P. Chuang, S. Ho, L. W. Smith, F. Sfigakis, M. Pepper, C. Chen, J. Fan, J. P. Griffiths, I. Farrer, H. E. Beere, G. A. C. Jones, D. A. Ritchie, and T. Chen, All-electric all-semiconductor spin field-effect transistors. Nature Nanotech 10, 35 (2015).

[9] S. Luo, M. Song, X. Li, Y. Zhang, J. Hong, X. Yang, X. Zou, N. Xu, and L. You, Reconfigurable skyrmion logic gates, Nano Lett. 18, 1180 (2018).

[10] D. Bhowmik, L. You, and S. Salahuddin, Spin Hall effect clocking of nanomagnetic logic without a magnetic field, Nat. Nanotech. 9, 59 (2014).

[11] C. Murapaka, P. Sethi, S. Goolaup, and W. S. Lew, Reconfigurable logic via gate-controlled domain wall trajectory in magnetic network structure. Sci. Rep. 6, 20130 (2016).

[12] J. A. Currivan-Incorvia, S. Siddiqui, S. Dutta, E. R. Evarts, J. Zhang, D. Bono, C. A. Ross, and M. A. Baldo, Logic circuit prototypes for three-terminal magnetic tunnel junctions with mobile domain walls, Nat. Commun. 7, 10275 (2016).

[13] Z. Luo, A. Hrabec, T. P. Dao, G. Sala, S. Finizio, J. Feng, S. Mayr, J. Raabe, P. Gambardella, and L. J. Heyderman, Current-driven magnetic domain-wall logic, Nature 579, 214 (2020).

[14] S. Manipatruni, D. E. Nikonov, C. Lin, T. A. Gosavi, H. Liu, B. Prasad, Y. Huang, E. Bonturim, R. Ramesh,

and I. A. Young, Scalable energy-efficient magnetoelectric spin-orbit logic, Nature 35, 565 (2019).

[15] Y. Chai, Y. Liang, C. Xiao, Y. Wang, B. Li, D. Jiang, P. Pal, Y. Tang, H. Chen, Y. Zhang, H. Bai, T. Xu, W. Jiang, W. Skowroński, Q. Zhang, L. Gu, J. Ma, P. Yu, J. Tang, Y. Lin, D. Yi, D. C. Ralph, C. Eom, H. Wu, and T. Nan, Voltage control of multiferroic magnon torque for reconfigurable logic-in-memory. Nat. Commun. 15, 5975 (2024).

[16] M. Yang, Y. Deng, Z. Wu, K. Cai, K. W. Edmonds, Y. Li, Y. Sheng, S. Wang, Y. Cui, J. Luo, Y. Ji, H. Zheng, and K. Wang, Spin logic devices via electric field controlled magnetization reversal by spin-orbit torque, IEEE Electron Device Lett. 40, 155 (2019).

[17] S. C. Baek, K. Park, D. Kil, Y. Jang, J. Park, K. Lee, and B. Park, Complementary logic operation based on electric-field controlled spin-orbit torques, Nat. Electron. 1, 398 (2018).

[18] M. Li, C. Li, X. Xu, M. Wang, Z. Zhu, K. Meng, B. He, G. Yu, Y. Hu, L. Peng, and Y. Jiang, An ultrathin flexible programmable spin logic device based on spin-orbit torque, Nano Lett. 23, 3818 (2023).

[19] R. Posti, A. Ravindran K, D. Tiwari, and D. Roy, Versatility of spin-logic and high-density multistate memory enabled by a single spin-orbit torque device, ACS Appl. Electron. Mater. 7, 3955 (2025).

[20] Y. Dong, T. Xu, H. Zhou, L. Cai, H. Wu, J. Tang, and W. Jiang, Electrically reconfigurable 3d spin-orbitronics, Adv. Funct. Mater. 31, 2007485 (2021).

[21] C. Wan, X. Zhang, Z. Yuan, C. Fang, W. Kong, Q. Zhang, H. Wu, U. Khan, and X. Han, Programmable spin logic based on spin hall effect in a single device. Adv. Electron. Mater. 3, 1600282 (2017).

[22] Y. Zhao, G. Yang, J. Shen, S. Gao, J. Zhang, J. Qi, H. Lyu, G. Yu, K. Jin, and S. Wang, Implementation of complete Boolean logic functions in single spin-orbit torque device. AIP Adv. 11, 015045 (2021).

[23] Z. Luo, Z. Lu, C. Xiong, T. Zhu, W. Wu, Q. Zhang, H. Wu, X. Zhang, and X. Zhang, Reconfigurable magnetic logic combined with nonvolatile memory writing, Adv. Mater. 29, 1605027 (2017).

[24] Y. Fan, X. Han, X. Zhao, Y. Dong, Y. Chen, L. Bai, S. Yan, and Y. Tian, Programmable spin-orbit torque multistate memory and spin logic cell, ACS Nano 16, 6878 (2022).

[25] X. Huang, Y. Zhao, X. Wang, F. Wang, L. Liu, H. Yang, W. Zhao, and S. Shi, Implementing versatile programmable logic functions using two magnetization switching types in a single device, Adv. Funct. Mater. 34, 2308219 (2024).

[26] T. Zhao, Z. Zheng, J. Wang, G. Zhou, L. Liu, C. Zhou, Q. Xie, L. Jia, R. Xiao, Q. Zhang, L. Ren, S. Shi, T. Zeng, Y. Gu, X. Xu, Y. Zhang, and J. Chen, Spin logic enabled by current vector adder, Nat. Commun. 16, 2988 (2025).

[27] Z. Zheng, Z. Zhang, X. Feng, K. Zhang, Y. Zhang, Y. He, L. Chen, K. Lin, Y. Zhang, P. K. Amiri, and W. Zhao, Anomalous thermal-assisted spin-orbit torque-induced magnetization switching for energy-efficient logic-in-memory, ACS Nano 16, 8264 (2022).

[28] X. Zhao, Y. Dong, W. Chen, X. Xie, L. Bai, Y. Chen, S. Kang, S. Yan, and Y. Tian, Purely electrical controllable complete spin logic in a single magnetic heterojunction, Adv. Funct. Mater. 31, 2105359 (2021).

[29] X. Wang, C. Wan, W. Kong, X. Zhang, Y. Xing, C. Fang, B. Tao, W. Yang, L. Huang, H. Wu, M. Irfan, and X. Han, Field-free programmable spin logics via chirality-reversible spin-orbit torque switching, Adv. Mater. 30, 1801318 (2018).

[30] J. Lin, S. Zhang, S. Li, Y. Xu, X. Li, W. Duan, J. Hou, C. Zhou, W. Zhan, Z. Guo, M. Song, X. Yang, Y. Tian, X. Zou, D. Feng, and L. You, Initialization-free programmable spin-logic gate in a single spin-orbit torque device, Engineering 51, 215 (2025)

[31] Q. Liu, L. Liu, G. Xing, and L. Zhu, Asymmetric magnetization switching and programmable complete Boolean logic enabled by long-range intralayer Dzyaloshinskii-Moriya interaction, Nat. Commun. 15, 2978

(2024).
[32] L. Zhu, L. Zhu, S. Shi, D. C. Ralph, and R. A. Buhrman, Energy-efficient ultrafast SOT-MRAMs based on low-resistivity spin Hall metal $Au_{0.25}Pt_{0.75}$, Adv. Electron. Mater. 6, 1901131 (2020).
[33] G. Han, X. Lin, Q. Liu, G. Gong, L. Zhu, Invalidation of the domain wall depinning model and current-induced switching angle shift analysis in $Pt_{75}Ti_{25}$/Ti/$Fe_{60}Co_{20}B_{20}$ heterostructure, Adv. Funct. Mater. 36, e23908 (2025).
[34] X. Yin, G. Han, G. Gong, J. Kang, C. Xiong, and L. Zhu, Physical origin of current-induced switching angle shift in magnetic heterostructures, Chin. Phys. Lett. 42, 11703 (2025).
[35] L. Zhu, L. Zhu, S. Shi, M. Sui, D. C. Ralph, and R. A. Buhrman, Enhancing spin-orbit torque by strong interfacial scattering from ultrathin insertion layers, Phys. Rev. Appl. 11, 061004 (2019).
[36] S. Jung, H. Lee, S. Myung, H. Kim, S. K. Yoon, S. Kwon, Y. Ju, M. Kim, W. Yi, S. Han, B. Kwon, B. Seo, K. Lee, G. Koh, K. Lee, Y. Song, C. Choi, D. Ham, and S. J. Kim, A crossbar array of magnetoresistive memory devices for in-memory computing. Nature 601, 211 (2022).

**Acknowledgement:**
**Funding:** This work was supported partly by the National Key Research and Development Program of China (2022YFA1204000), by the Beijing Natural Science Foundation (Z230006), and by the National Natural Science Foundation of China (12304155, 12274405).

**Author contributions:** L. Z. conceived the project, Q. L. fabricated the samples and performed the measurements, and L. Z. and Q. L. wrote the manuscript.

**Competing interests:** The authors declare no competing interests.

**Data and Materials Availability:** All data are present in the paper.